\documentclass[preprint,showpacs,preprintnumbers,amsmath,amssymb,superscriptaddress]{revtex4}
\usepackage[dvipdfmx]{graphicx}% Include figure files
\usepackage[dvipdfmx]{color} %display png file
\usepackage{dcolumn}% Align table columns on decimal point
\usepackage{bm}% bold math
\usepackage{braket}%braket
\usepackage{amsmath}
\usepackage{subfigure}
\usepackage{color}
\usepackage[export]{adjustbox}

\begin{document}

%\newgeometry{left=2.5cm,bottom=1cm}
%\setstretch{2.4}
%\preprint{APS/123-QED}
\title{{Effects of hole-boring and relativistic transparency on particle acceleration in overdense plasma irradiated by short multi-PW laser pulses
}}
\author{Masahiro Yano}
\thanks{myano@ef.eie.eng.osaka-u.ac.jp}
%\providecommand{\keywords}[1]{\textbf{\textit{Index terms---}} #1}
%\email{myano@ef.eie.eng.osaka-u.ac.jp, +81-06-6879-7803}
\affiliation{Graduate School of Engineering, Osaka University, 2-1, Yamadaoka, Suita, Osaka 565-0871, Japan}
\author{Alexei Zhidkov}
\affiliation{Graduate School of Engineering, Osaka University, 2-1, Yamadaoka, Suita, Osaka 565-0871, Japan}
\author{James Koga}
\affiliation{Kansai Photon Science Institute, National Institutes for Quantum and Radiological Science and Technology, Kizugawa, Kyoto 619-0215, Japan}
\author{Tomonao Hosokai}
\affiliation{Department of Beam Physics, Division of Advanced Materials and Beam Science, The Institute of Scientific and Industrial Research, Osaka University, 8-1 Mihogaoka, Ibaraki, Osaka 567-0047, Japan}
\affiliation{Laser Accelerator R\&D Team, RIKEN SPring-8 Center, 1-1-1, Kouto, Sayo-cho, Sayo-gun, Hyogo 679-5148, Japan}
\author{Ryosuke Kodama}
\affiliation{Graduate School of Engineering, Osaka University, 2-1, Yamadaoka, Suita, Osaka 565-0871, Japan}
\affiliation{Institute of Laser Engineering, Osaka University, 2-1, Yamadaoka, Suita, Osaka 565-0871, Japan}
%\date{\today}
%{\let\newpage\relax\maketitle}
%\pacs{52.57.Kk, 52.40.Mj, 52.50.Gj}% PACS,

\begin{abstract}

Propagation of short and ultra-intense laser pulses in a semi-infinite space of overdense hydrogen plasma is analyzed via fully-relativistic, real geometry particle-in-cell (PIC) simulations including radiation friction. The relativistic transparency and hole-boring regimes are found to be sensitive to the transverse plasma field, backward light reflection, and laser pulse filamentation. For laser intensities approaching $I\sim10^{24}$ W/cm$^2$ the direct laser acceleration of protons, along with ion Coulomb explosion, results in their injection into the acceleration phase of the compressed electron wave at the front of the laser pulses. The protons are observed to be accelerated up to 10-20 GeV with densities around a few times the critical density. The effect strongly depends on initial density and laser intensity disappearing with initial density increase and intensity decrease.
\end{abstract}

\maketitle
\newpage
		
Recently the power of femtosecond laser pulses has exceeded the petawatt level and is continuously increasing\cite{Danson}. Intensities over 10$^{23}$ W/cm$^2$ ($a_0\sim300$, where $a_0$ is the normalized laser field \cite{Esarey,Mourou}) at the laser focus have already been reached\cite{Shin}. Several upcoming projects \cite{Ur,Miyanaga,Crane,Batani,Maywar} make possible laser intensities exceeding $I\sim10^{24}$ W/cm$^2$ or $a_0\sim10^3$. That would be a qualitative step forward in high field physics, because the parameter $ma_0/M$, where $M$ is the proton mass, becomes of the order of unity. Such powerful laser pulses can directly accelerate heavy particles up to relativistic energies. Moreover, the effect of relativistic transparency theoretically allows laser pulse propagation through even solid density plasmas. The latter, along with the generation of energetic particles, has been of particular interest.

Specific to the interaction of powerful laser pulses and overdense plasmas is the relativistic effects in the plasma. A multi-PW laser pulse has a normalized vector potential $a_0=eE/mc\omega$ [where $E$ is the amplitude of the laser field and $\omega$ is the laser pulse frequency] far exceeding unity. Therefore, the group velocity of a laser pulse in plasma $v_G=\sqrt{1-4\pi e^2 N_e/m\omega^2 a_0}$ \cite{Esarey,Mourou} remains positive for electron densities which are $a_0$ times higher than the critical density $N_{cr}=m\omega^2/4\pi e^2$. For Ti-Sph laser pulses and the laser intensity $I=10^{24}$ W/cm$^2$, the maximal density is $N_{emax}=2\times10^{24}$ cm$^{-3}$. However, the equation for the group velocity is correct only for the electron figure-eight motion which could be restricted by the transverse components of the plasma field. Additionally, the hole-boring process \cite{Iwata1,Tabak,Kodama,Levy,Wilks1,Pukhov,Sentoku,Naumova,Ping,Weng,Lei} can also allow high intensity laser pulses to propagate through a spatially semi-infinite overdense plasma. Due to the laser ponderomotive force, $F_p=-\frac{1}{4}  \frac{mc^2}{\gamma} \nabla a_0^2$ \cite{Esarey,Mourou} electrons are pushed at the front of laser pulses while the main body of the pulses propagates in a much lower density plasma. Again, such processes as pulse filamentation and other instabilities can drastically diminish this force and finally stop the hole-boring process. How far powerful laser pulses can propagate in a spatially semi-infinite overdense plasma with different density and the domains of validity of such a process have yet to be fully understood.

Another important feature of the interaction of ultra-high intensity laser pulses with plasma is the possibility of direct acceleration of heavy ions by laser pulses. The general solution of particle motion in the plane wave gives \cite{Esarey,Mourou,Yogo,Daido,Wilks2,Snavely,Esirkepov,Yin,Fiuza,Haberberger,Hegelich,Palaniyappan}:
	\begin{equation*}
\gamma=\sqrt{(M/m)^2+p_X^2+p_Y^2}; \;\; \gamma-p_X=A; \;\; p_Y=a_0 \cos(\omega t - \frac{\omega x}{c})
	\end{equation*}
where $M$ is the ion mass, $A$ is a constant and $p$ is normalized by $mc$, the particle momentum. If the ion is initially at rest, $A=M/m$ and $p_X=mp_Y^2/2M$. For $a_0> M/m$ the energy of ion becomes relativistic even in a vacuum. In a relativistically transparent plasma, this effect may occur as a self-injection process for ions at the front of a laser pulse.

 Additionally, a reflection of laser light in relativistically transparent plasma has a particular interest. First of all, in the relativistic transparency case, Raman scattering may occur, because the plasma frequency $\omega_p^\prime=\omega_p / \sqrt{a_0}$ can become less than the laser frequency. However, its detection is very difficult because its frequency  $\omega=\omega_L\pm \omega_p$ is too close to the laser frequency. More interesting is the feature of reflected light. The reflecting boundary moves with a speed close to the speed of light. Therefore, a strong frequency downshifts should occur with a minimal frequency cut corresponding to the maximal velocity of reflecting boundary.

In this letter, we investigate numerically the dynamics of ultra-intense laser pulses and the generation of energetic particles in spatially semi-infinite, overdense hydrogen plasma. Multidimensional PIC simulations including fully relativistic motions for electrons and ions as well as the classical radiation friction force are performed with high spatial $\lambda/200$ in 2D and maximal $\lambda/50$ in 3D using the code FPlaser \cite{Zhidkov,Yano} with the moving window technique. Limited spatial resolution for 3D simulations constrains the range of plasma density and plasma length.

Linearly polarized laser pulses with wavelength $\lambda = 1$ $\mu m$ and duration $\tau = 10$ fs propagate in the -x (longitudinal) direction from the right to left in a pre-ionized semi-infinite plasma. The laser pulse intensity is varied from 10$^{23}$ to 10$^{24}$ W/cm$^2$, corresponding to $a_0$ from $\sim10^2$ to $\sim10^3$. The pulse intensities 10$^{23}$ to 10$^{24}$ W/cm$^2$ at the laser focus correspond to energies from $\sim$0.75 kJ to $\sim$7.5 kJ and powers of $\sim$75 to $\sim$750 PW. Laser intensities exceeding $I = 10^{22}$ W/cm$^2$ have already been achieved \cite{Shin} and the maximum laser intensity is approaching $I=10^{24}$ W/cm$^2$ as reported in \cite{Ur,Miyanaga,Crane,Batani,Maywar}, which makes this range of laser intensities interesting for investigation. The initial conditions for the transverse components of the fields are taken as the well-known solution of the para-axial equations \cite{Zhidkov,Yano} with the waist $w_0=5$ $\mu$m and corresponding Rayleigh length $\sim$75 $\mu$m. In simulations, the size of the moving window is (100 $\mu$m) $\times$ (100 $\mu$m) $\times$ (110 $\mu$m) to resolve high plasma frequency. The density of the uniform plasma is a parameter ranging from $N_e = 2N_{cr}$ to $100N_{cr}$. Such plasma can be produced in a mixture of high $Z$ gases with hydrogen or in exploding wires \cite{Vijayan,Iwata2}. However, here we consider pure hydrogen plasma to minimize the number of physical processes. The linear density ramp in the front of the plasma has a length $L = 10 \mu$m. Absorbing boundary conditions are used in the code \cite{Zhidkov,Yano}.  

As a result of the simulations, we found that the laser pulses have a finite penetration depth, which depends on the plasma density and initial laser pulse intensity. Typical results are presented in Fig.1 and Fig.2. In Fig.1 the dependences of the penetration depth, $L_D$, for laser pulses with different intensity on the plasma density are shown. The penetration depth is calculated as the distance from the plasma boundary where the laser pulse vanishes or starts moving backward. Even for double critical density and $a_0\sim10^3$ the penetration depth is finite; for $I=10^{24}$ W/cm$^2$, $L_D\sim 300$ $\mu$m even though $N_e/N_{cr}a_0$ remains far less than unity. For $N_e=100N_{cr}$ the propagating laser pulse scatters from the boundary. The processes limiting the propagation length can be seen in Fig.2 where the laser field is given at the stopping point. For lower density, the main reason is the filamentation instability \cite{Esarey,Mourou} as seen in Fig 2a,b, along with the Raman scattering decreasing the pulse intensity. This underlines the role of the hole-boring process in the propagation of laser pulses. Filamented laser pulses cannot form a proper channel for guiding and plasma electrons cannot acquire the energy necessary for the relativistic transparency. With density increase, the transverse plasma field does not allow the 8-figure motion of electrons in the laser field and, as a result, the laser pulse does not penetrate in plasma essentially even though the plasma is still theoretically relativistically transparent. One can see strong scattered light from the plasma surface. With decreasing laser intensity all of the above effects occur earlier and, therefore, $L_D$ becomes smaller.

During the laser pulse propagation, a plasma cavity having a lower density than the surrounding plasma forms in the pulse wake as seen in Fig.3. This is the result of electron evacuation from the laser axis and further ion Coulomb explosion. Therefore, most of the ions acquire large transverse momenta. The cavity shape is dominated by the dynamics of energetic electrons. In Fig.4a,b the effect of the radiation friction force is illustrated by the shape of cavities with and without it. One can see that the radiation friction force stabilizes the cavity shape. Since the electron density in the cavity is low, the electron acceleration occurs in an underdense plasma driven by the wake field. The results of electron acceleration can be seen in Fig.5a,b for different intensities. As expected, higher energy electrons appear in low, $N_e=2N_{cr}$, plasma. The maximum electron energy decreases with increasing density reflecting the shortening of the efficient acceleration length in the wake field. The maximum energy increases as $a_0$ for lower density and increases more slowly at higher density. 

The results for ion acceleration presented in Fig.6a,b, show a completely different picture from the case of electron acceleration. The ion energy distribution function at the lower laser intensity, $I=10^{23}$ W/cm$^2$, is shown in Fig.6a. An increase of plasma density results in slower phase velocity of the plasma wake wave. Since background ions are initially accelerated behind the laser pulse due to the Coulomb explosion, efficient injection of these ions into the plasma wake wave can occur when their velocities match the plasma wake phase velocity. At $N_e=10N_{cr}$ the maximal energy is near 1 GeV. At low density, the maximal ion energy is rather low of the order of 100 MeV. For plasma with density $N_e=100N_{cr}$, the maximum ion energy is also low. The process of ion acceleration drastically changes for $I=10^{24}$ W/cm$^2$ where direct ion acceleration becomes more dominant. For lower plasma density, $N_e=2N_{cr}$, the direct ion acceleration by the laser pulse serves as an injector of ions into the acceleration phase of the first plasma wave bucket at the front of the laser pulse. Then, a part of such injected ions is accelerated up to 20 GeV. These ions were positioned at the front of the laser pulses in our numerical simulations. With increasing density the direct ion acceleration becomes less efficient for injection as compared to the Coulomb explosion and, therefore, the maximum energy rapidly goes down. The radiation friction force does not affect this mechanism of ion injection and acceleration. 

Spectra of plasma radiation, in the range of our spectral resolution, calculated via the wave vectors at the times of the laser pulse stopping for different intensities and plasma densities are shown in Fig.7a,b. Forwardly and backwardly propagating lights cannot be separated in detail in such a calculation. However, overall, we can attribute the lower frequency radiation to the backward radiation and higher frequencies to the forward radiation. Since the calculations are done for the fields with increasing laser intensity the amplitude increases proportionally to $a_0$. One can see rather complicated spectra in the lower frequency range. Due to the strong plasma non-uniformity, the spectra are not so clean. The maximal frequency downshifts, according to theory, should be $\lambda_D=4\lambda N_c a_0/N_e$. For maximal laser intensities, this wavelength is too long and lies out of our special resolution. However, in the stop point, the $a_0$ becomes lower and spectra reflect velocity distribution in the plasma wave. One can see several peaks in the long wavelength spectrum we attribute them to backward light reradiated from the walls of the plasma channel. At the stop points, we also observe strong radiation in the higher frequency range. We cannot precisely determine the source of radiation. However, the appearance of this radiation mainly in the vicinity of the stop point indicates that this is radiation from braking electrons in the first plasma wave bucket. Limitations in the resolution of the simulations to $\lambda/150$ did not permit studying such radiation in detail.

In conclusion, we have observed a novel effect of ion acceleration by multi-PW laser pulses in relativistically underdense spatially semi-infinite plasma via fully relativistic PIC simulations including electron and ion motion along with the radiation damping. This acceleration occurs in plasmas with electron densities around 2-10 critical density due to the direct proton acceleration by laser pulses with $a_0\sim 10^3$ resulting in proton injection into the acceleration phase of the plasma wave at the front of the laser pulses. Protons with energies up to 20 GeV have been observed for 10 fs laser pulses with intensity $I=10^{24}$ W/cm$^2$ irradiating overdense plasma with $N_e=2N_{cr}$. Such a plasma can be produced by a mixture of a high $Z$ gas and hydrogen. The proton energy rapidly decreases with increasing plasma density. For lower laser intensity the effect vanishes since direct proton acceleration becomes impossible. The behavior of plasma electrons does not show any essential difference from the conventional physical picture and recent experiments with lower laser intensities: energetic electrons form a broad Maxwell-like distribution with an effective temperature proportional to $a_0$. 

Limitation of the laser pulse propagation in a theoretically relativistically transparent spatially semi-infinite plasma has been observed over a wide range of laser intensities and plasma densities. Laser pulse filamentation, backward Raman scattering, and the transverse plasma field break the transparency condition $N_e/N_{cr}a_0<1$.

Forward and backward radiation can be an important tool for studying the interaction of ultra-high power laser pulses with plasma. In the present simulations, analysis of backward radiation spectra has shown features can be explained by Doppler frequency downshift during the light reflection from the moving cavity boundary. However, a full analysis of backward radiation spectra requires a special spatial resolution since the maximal frequency cut-off lies in the THz region. The short wavelength forward radiation can be explained via the stopping process when electrons, directly accelerated by the laser pulse, decelerate. Such radiation does not occur in the earlier stage of laser pulse propagation. In spite of the complicated character of plasma radiation spectra, their detection should add some clarity to the interaction of high power laser pulses with relativistically-underdense, spatially semi-infinite plasma.

		\section*{Acknowledgement}

This work was supported and funded by the ImPACT Program of the Council for Science, Technology, and Innovation (Cabinet Office, Government of Japan). Part of this work was also supported by the JST-MIRAI Program Grant No.JPMJMI17A1. This work was (partially) achieved through the use of large-scale computer systems at the Cybermedia Center, Osaka University.

\clearpage
\begin{figure}[t]
%\iffigure
\begin{center}
\includegraphics[width=0.9\hsize]{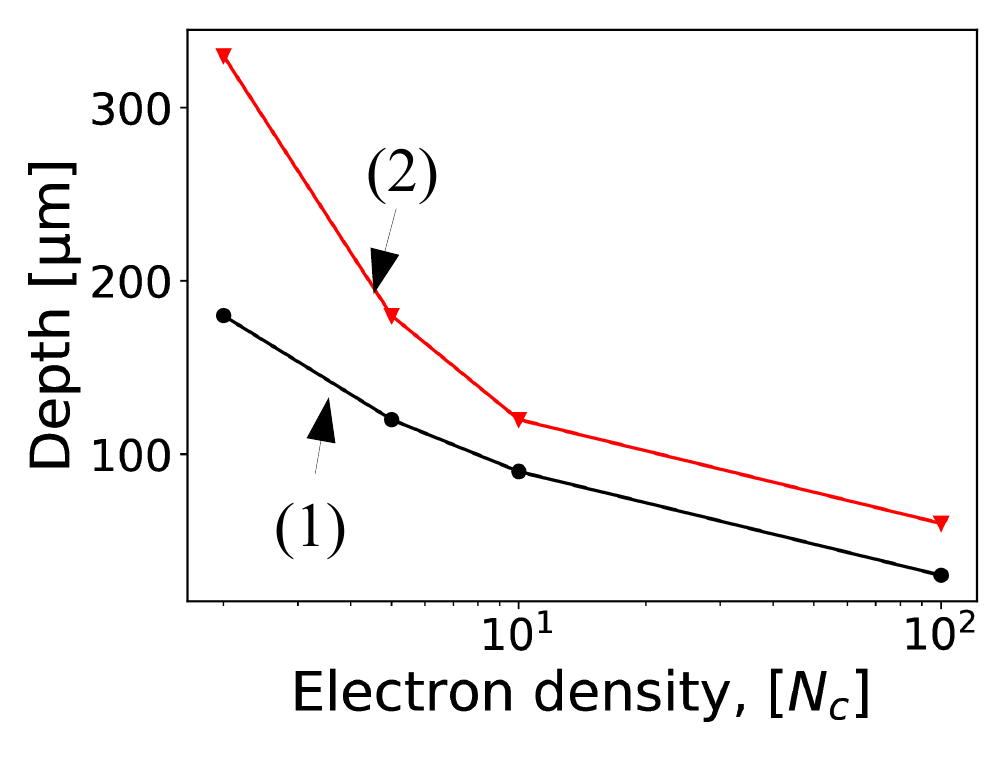}
\caption{ Plot of penetration length for intensities (1) $I=10^{23}$  W/cm$^2$, and (2) $I=10^{24}$  W/cm$^2$. The penetration length is shorter for higher electron density and lower laser intensity.}
\end{center}
\end{figure}

\clearpage
\begin{figure}[t]
%\iffigure
\begin{center}
\includegraphics[width=0.9\hsize]{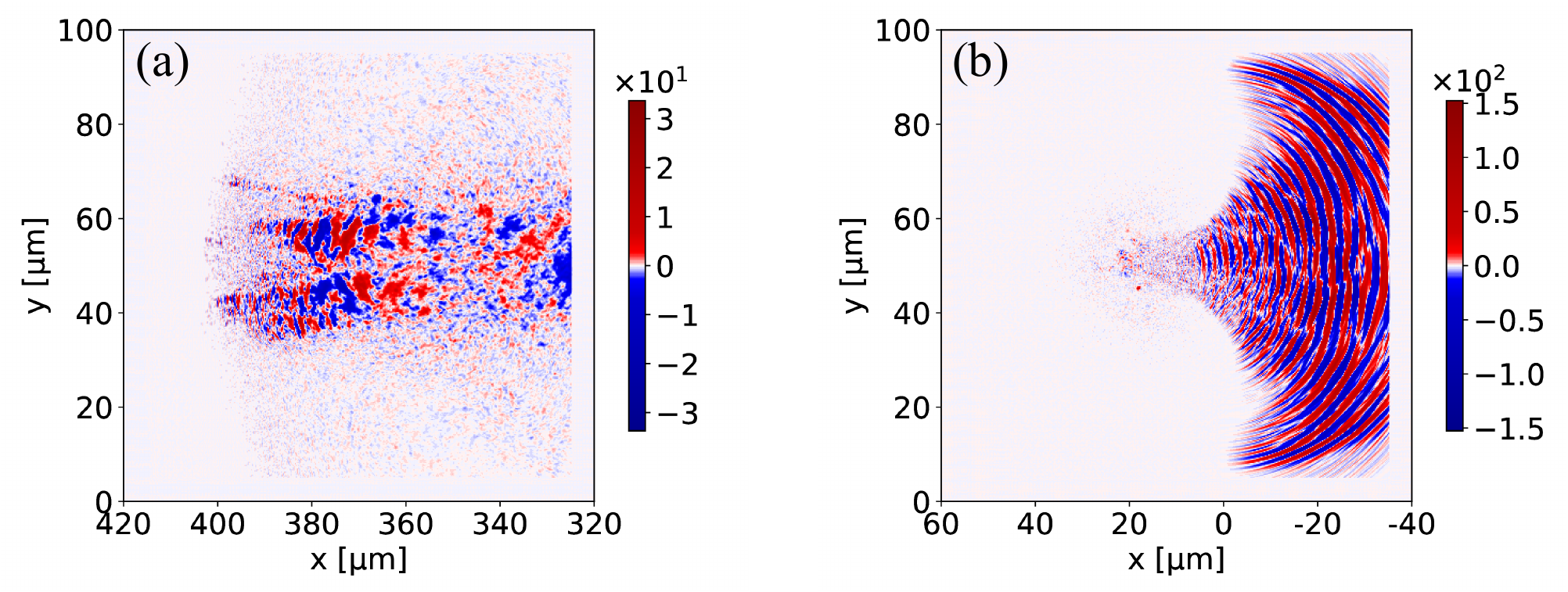}
\caption{$xy$ cross-section of a laser pulse at the moment when the pulse propagation stops for (a) initial plasma density 2$N_{cr}$, intensity $I=10^{24}$  W/cm$^2$, the time t=1300fs, (b) initial plasma density 100$N{cr}$, intensity $I=10^{24}$  W/cm$^2$, the time t=200fs. }
\end{center}
\end{figure}
%\fi

\clearpage
\begin{figure}[t]
%\iffigure
\begin{center}
\includegraphics[width=0.5\hsize]{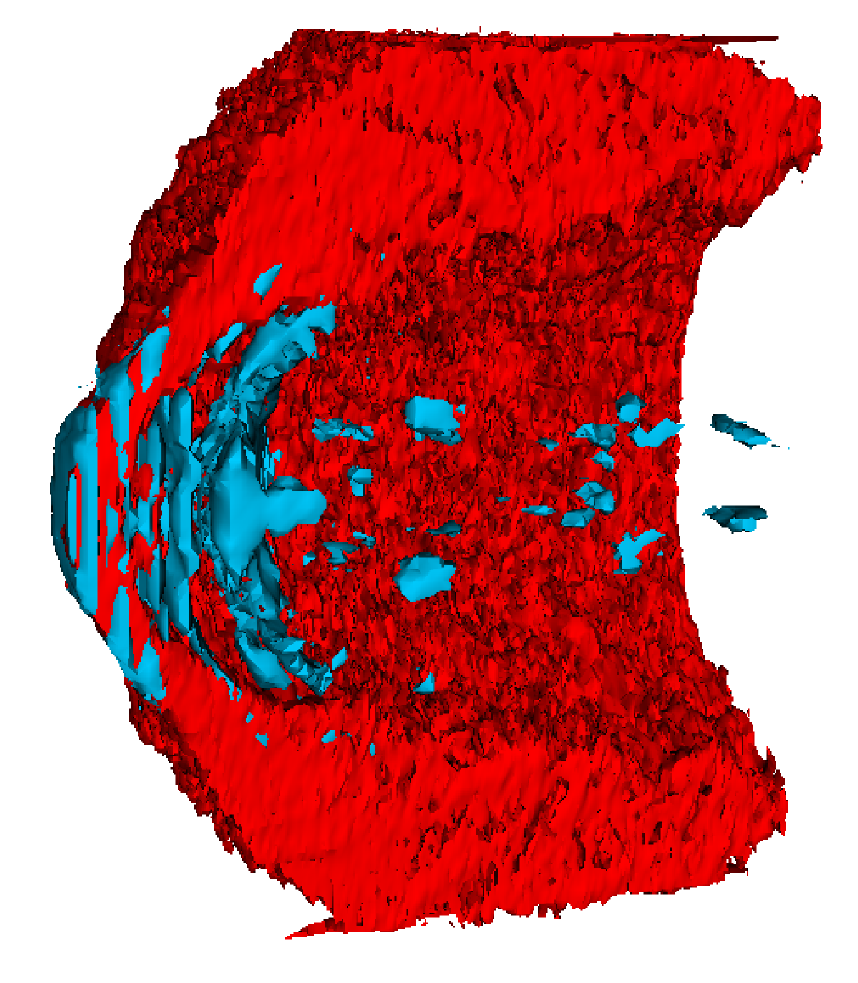}
\end{center}
%\fi
\caption{3D plot of a laser pulse and ion density where $z\ge$ 15 $\mu$m for initial plasma density 5$N_{cr}$, intensity $I=10^{24}$  W/cm$^2$, and the time t=300fs. Red shows ions. Blue shows a laser pulse. The laser pulse propagates from upper right to lower left.}
\label{fig:fig2}
\end{figure}

\clearpage
\begin{figure}[t]
%\iffigure
\begin{center}
\includegraphics[width=\hsize]{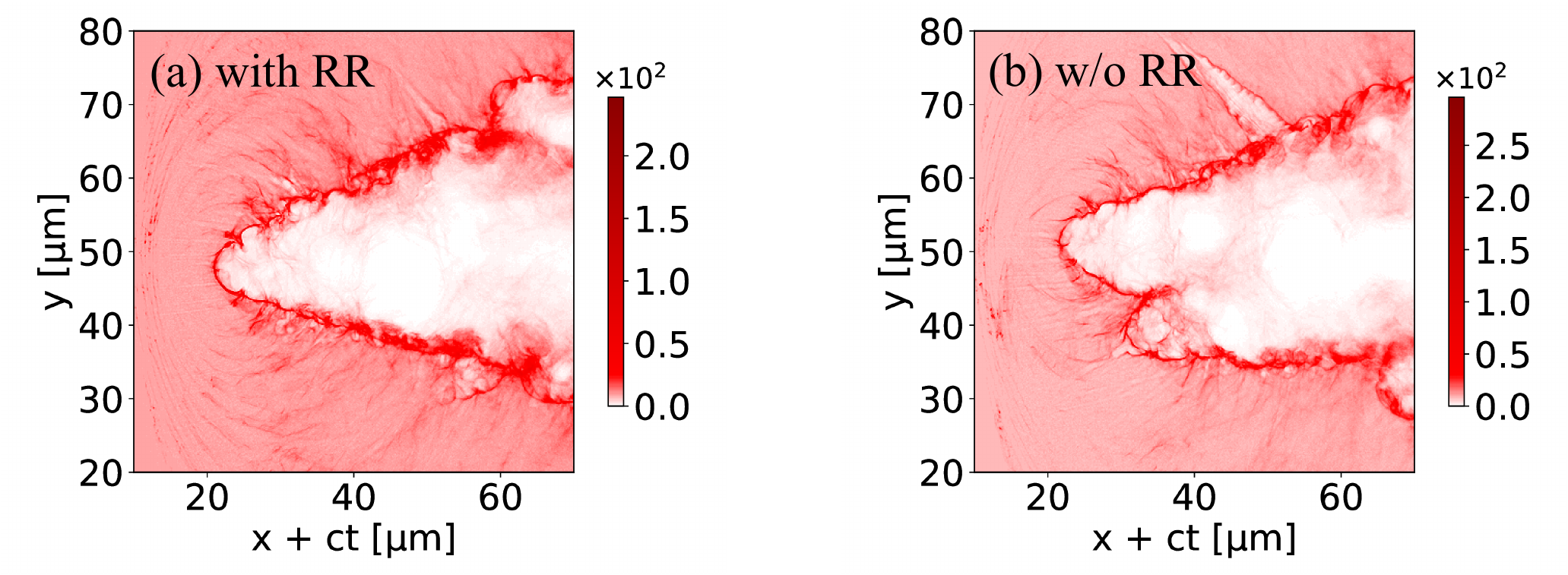}
\end{center}
%\fi
\caption{$xy$ cross-section of ion density at the moment when the pulse propagation stops for the initial plasma density 10$N_{cr}$, intensity $I=10^{24}$  W/cm$^2$, the time t=300fs, (a) with radiation reaction force and (b) without radiation reaction force.}
\label{fig:fig3}
\end{figure}

\clearpage
\begin{figure}[t]
%\iffigure
\begin{center}
\includegraphics[width=\hsize]{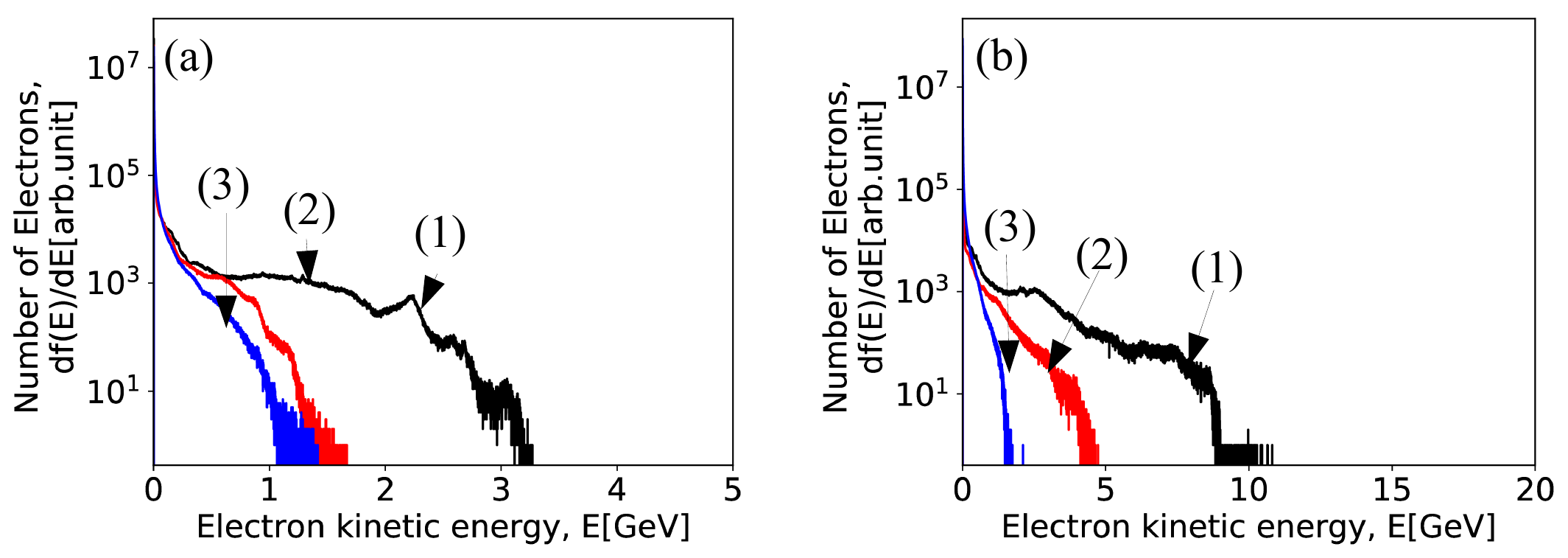}
\end{center}
%\fi
\caption{Electron energy spectrum for (a) intensity $I=10^{23}$  W/cm$^2$, (b) intensity $I=10^{24}$  W/cm$^2$. Lines (1), (2), and (3) indicate the case for initial densities of 2$N_{cr}$, 5$N_{cr}$, and 10$N_{cr}$, respectively.}
\label{fig:fig4}
\end{figure}

\clearpage
\begin{figure}[t]
%\iffigure
\begin{center}
\includegraphics[width=\hsize]{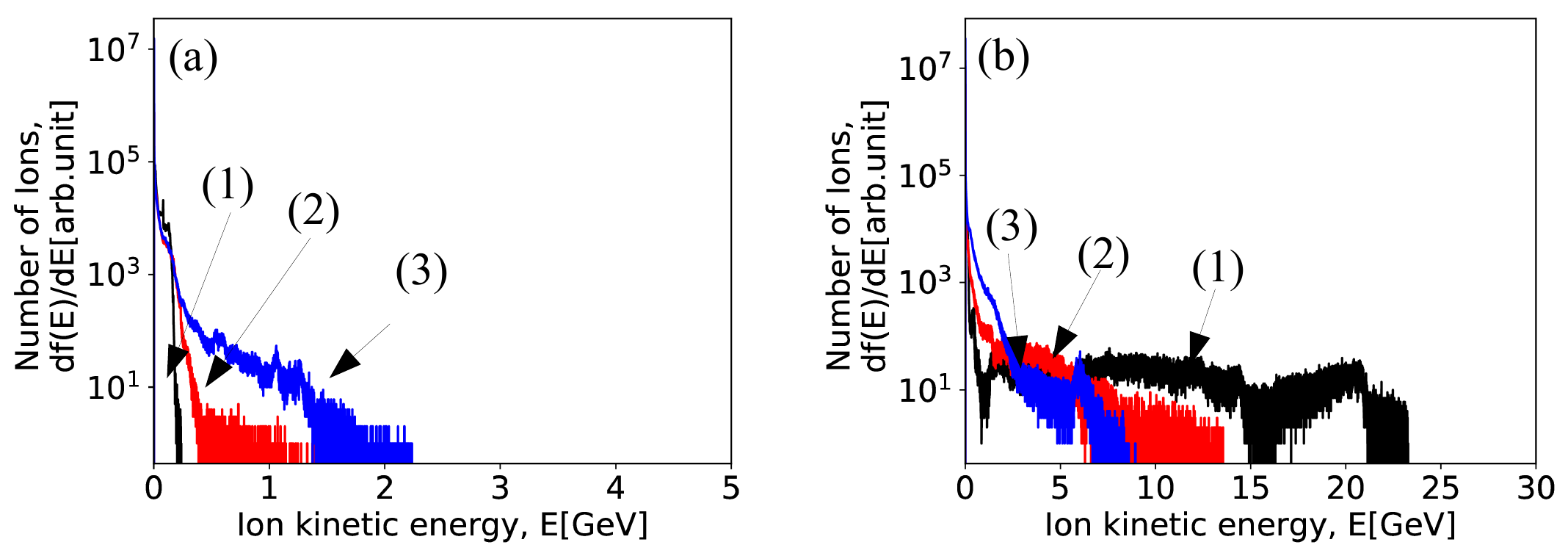}
\end{center}
%\fi
\caption{Ion energy spectrum for (a) intensity $I=10^{23}$  W/cm$^2$, (b) intensity $I=10^{24}$  W/cm$^2$. Lines (1), (2), and (3) indicate the case for initial densities of 2$N_{cr}$, 5$N_{cr}$, and 10$N_{cr}$, respectively.
}
\label{fig:fig5}
\end{figure}

\clearpage
\begin{figure}[t]
%\iffigure
\begin{center}
\includegraphics[width=\hsize]{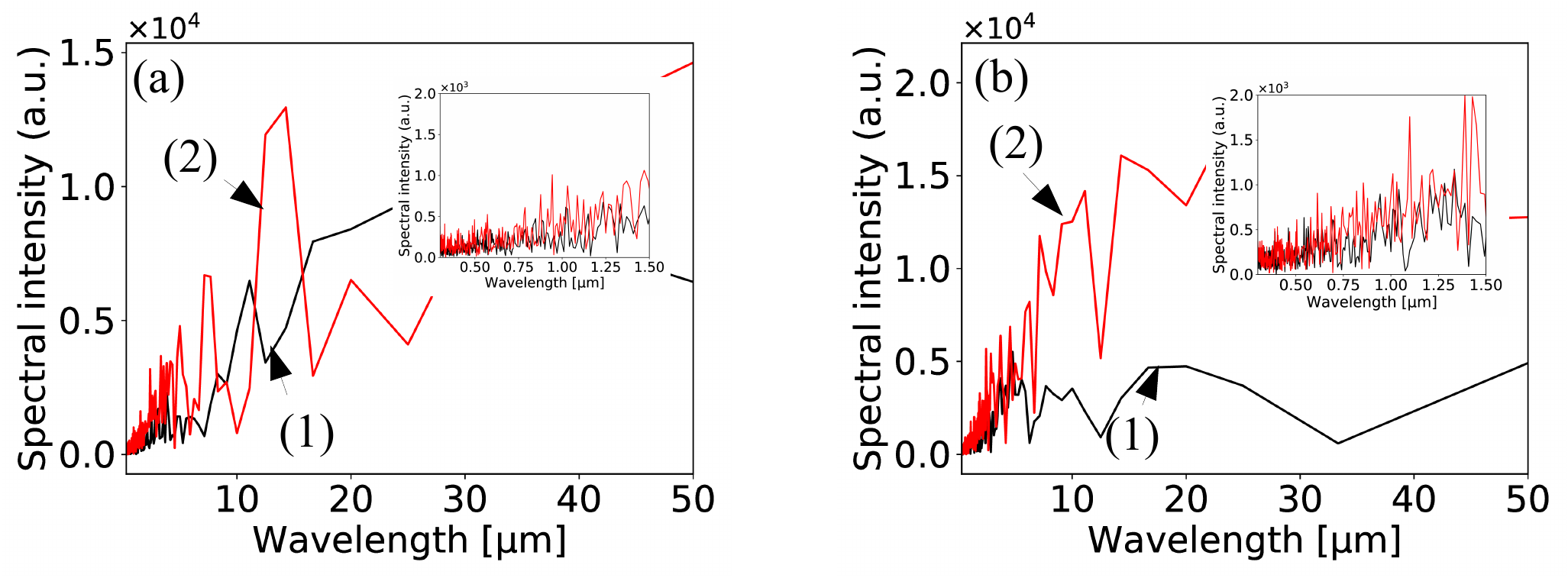}
\end{center}
%\fi
\caption{Spectral calculations of laser pulses when its propagation stops (a) for initial plasma densities of 5$N_{cr}$ and (b) 10$N_{cr}$. (1) The black line indicates the case for intensity $I=10^{23}$  W/cm$^2$ and (2) the red line indicates the case for $I=10^{24}$  W/cm$^2$. Inserts show the plots of shorter wavelengths $<$ 1.5 $\mu$m.
}
\label{fig:fig6}
\end{figure}

\end{document}